\documentclass{ar-1col-S2O}
\usepackage[numbers]{natbib}
\usepackage{times}
\usepackage{amsmath}
\usepackage{float}
\usepackage{xcolor}
\usepackage{graphicx}
\usepackage{amssymb}
\usepackage{url,hyperref}
\usepackage{comment}

\newcommand{\mlm}[1]{\textcolor{black}{#1}}
\newcommand{\ka}[1]{\textcolor{black}{#1}}
\newcommand{\tah}[1]{\textcolor{black}{#1}}

\jname{Xxxx. Xxx. Xxx. Xxx.}
\jvol{AA}
\jyear{YYYY}
\doi{10.1146/((please add article doi))}

\begin{document}

\markboth{Aspinwall et al.}{Rigidity in biomechanical networks}


\title{Rigidity and mechanical response in biological structures}
\author{Kelly Aspinwall, Tyler Hain, M. Lisa Manning
\affil{Physics Department and BioInspired Institute, Syracuse University, Syracuse NY 13244 USA}}

\begin{abstract}
Rigidity is an emergent property of materials -- it is not a feature of individual components that comprise the structure, but instead arises from interactions between many constituent parts. Recently, it has been recognized that floppy-rigid or fluid-solid transitions are harnessed by biological systems at all scales to drive form and function. This review focuses on the different mechanisms that can drive emergent rigidity transitions in biomechanical networks, and describes how they arise in mathematical formalisms and how they are observed in practice in experiments. The goal is to aid researchers in identifying mechanisms governing rigidity in their biological systems of interest, highlight mechanical features that are universal across different systems, and help drive new scientific hypotheses for observed mechanical phenomena in biology. Looking forward, we also discuss how biological systems might tune themselves towards or away from such transitions over developmental or evolutionary timescales.
\end{abstract}

\maketitle

\tableofcontents

\section{Introduction}
Physicist Phil Anderson writes in his book on condensed matter, ``We are so accustomed to the rigidity of solid bodies -- the idea, for instance, that when we move one end of a ruler, the other end moves the same distance \ldots that we don’t accept its almost miraculous nature, that it is an `emergent property' not contained in the simple laws of physics, although it is a consequence of them."~\cite{anderson2018basic} 

\mlm{When a material is not rigid, it is called \emph{floppy}. In some materials where the connectivity or topology is allowed to change, a floppy material can fluidize, or undergo large deformations at zero cost to its structural energy.} Originally, theories to predict the emergence of rigidity were developed to aid in the engineering of structures like bridges and buildings, and in the understanding of crystalline solid state materials. In addition to these applications in non-living systems, rigidity transitions and the associated changes to emergent mechanics occur in biological systems at all scales, and recent work has suggested that they are utilized by organisms to drive form and function. Moreover, breakdowns in the spatiotemporal control of rigidity transitions lead to disease~\cite{KasCancerJamming,oswald_jamming_2017,MongeraCampas,wang_anisotropy_2020}. 

\begin{figure}
\includegraphics[scale= 0.5]{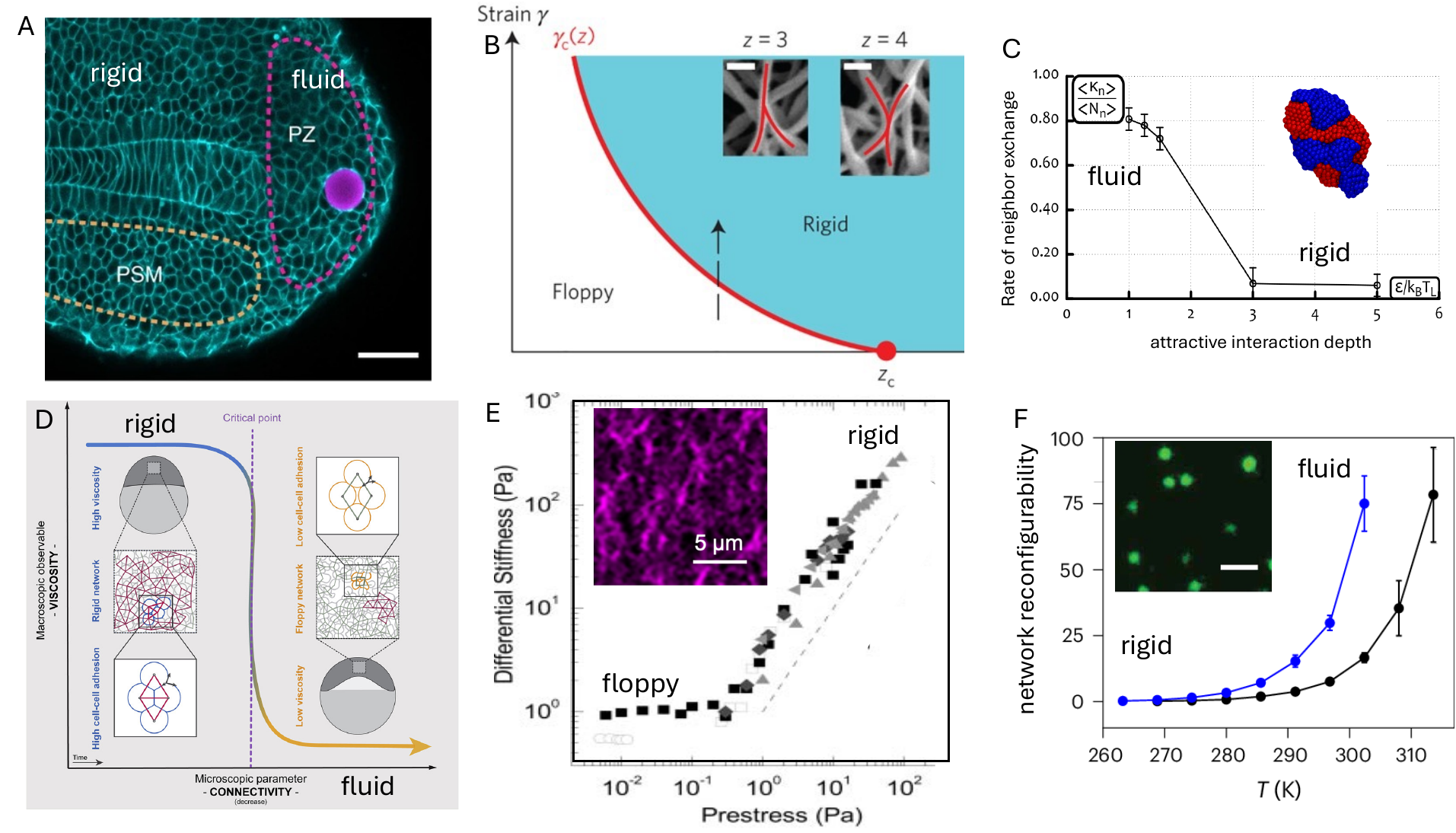}
\caption{{\bf Rigid-floppy transitions in biological mechanical networks.} (A) {\bf Zebrafish tailbud tissue.} 2D slice from a microscopy image of cell membranes in the zebrafish tailbud (cyan) with an embedded ferrofluid droplet(magenta)~\cite{serwane_vivo_2017}. Fluid-like and solid regions identified from viscosity data in~\cite{mongera_fluid--solid_2018}. (B) {\bf ECM networks.} Schematic of strain-stiffening behavior in a reconstituted collagen fiber networks as a function of applied strain $\gamma$ and connectivity $z$.  Inset: Images of collagen networks with connectivities of $z=3$ and $z=4$~\cite{sharma_strain-controlled_2016}. (C) {\bf Chromatin dynamics} in a polymer model for chromatin organization. As the attractive interaction between different zones in the chromatin increases, the rate of neighbor exchanges drops from a high value (fluid-like dynamics) to zero (solid-like dynamics)~\cite{michieletto2016polymer}.(D) {\bf Zebrafish epiboly.} Schematic diagram of viscosity vs. connectivity for zebrafish tissue during epiboly, showing that fluidity occurs after cell connectivity is decreased past a critical point. Insets - Diagrams of the cell-cell contact network model for a tissue of nonconfluent cells at high and low connectivities~\cite{petridou_rigidity_2021} (E) {\bf Cytoskeletal networks.} Strain stiffening behavior in reconstituted actin networks: measured stiffness as a function of the strain-induced prestress~\cite{gardel2006prestressed}. Inset: Confocal fluorescence image of reconstituted network of F‐actin~\cite{wu2022vimentin} {\bf Protein condensates.} At high temperatures, the network reconfigurability is high and the condensate is fluid-like, while at lower temperatures the reconfigurability is low and the condensate is rigid. A mutated protein (blue) remains fluid at lower temperatures compared to the standard one(black). Inset: image of protein condensate droplets that have become solid-like after aging. Scale bar 20 $\mu m$~\cite{shen2023liquid}. }
\label{fig:examples}
\end{figure}

For example, it has been shown that elongation of the zebrafish body axis during embryonic development depends on a carefully regulated fluid-solid transition in the tailbud tissue (Fig~\ref{fig:examples}A). Similarly, a fluid-solid transition at an earlier stage of zebrafish development is necessary to facilitate flows \mlm{(i.e. large-scale deformations)} of the tissue around the yolk that establish the anterior-posterior axis (Fig~\ref{fig:examples}D). Fluid-solid transitions have been observed in cultured tissues too~\cite{Angelini, devany_cell_2021}, and a delayed fluid-solid transition has been seen in cultured human lung tissues from patients with asthma~\cite{park_unjamming_2015, bi_density-independent_2015}. Fluid-solid transitions also help govern tissue patterning and compartmentalization in systems with stem-cell niches, such as the stratified epithelium~\cite{miroshnikova2018adhesion}. Changes to tissue fluidity are associated with tissue malignancy in cancer cell lines and primary tissues~\cite{grosser2021cell}, and fluid-solid transitions have been implicated in cancer metastasis~\cite{gottheil2023state}.

At smaller scales, the networks that comprise the cellular cytoskeleton (Fig~\ref{fig:examples}E) and the extracellular matrix (ECM, Fig~\ref{fig:examples}B) can also transition between floppy and rigid~\cite{licup_stress_2015, sharma_strain-controlled_2016}. \ka{In these networks, the topology of the network does not change as easily as in cellular structures and so the material is floppy (i.e. can be perturbed by small displacements with zero energy cost) but not fluid (i.e. does not permit large-scale structural changes or rearrangements.)} Inside the cell, the concentration of cytoskeletal components (such as actin or tubulin), the presence of crosslinking or branching proteins (like Arp2/3), and the activity of specialized crosslinking motor proteins (like myosin), all impact the structure and geometry of the network, which in turn control its rigidity and mechanics~\cite{gardel2008mechanical}. Similar components, such as cells that exert tension on nodes~\cite{Shenoy2}, interpenetrating soft networks that constrain stiff fibers~\cite{wyse2022structural}, and tension-sensitive severing proteins~\cite{Discher} control rigidity in extracellular networks.

At even smaller scales, the cytosol itself is so crowded that it can become glassy and solid-like in response to external cues~\cite{dix2008crowding}. In addition, mutations or aging in intrinsically disordered proteins can cause them to become stuck in a rigid phase instead of undergoing liquid-liquid phase separation~\cite{jawerth2020protein, alshareedah2024sequence, shen2023liquid} (Fig~\ref{fig:examples}F), potentially associated with disease states~\cite{alberti2021biomolecular}. Dense packings that generate solid-like restrictions in movement may also be important for chromatin dynamics and regulation~\cite{michieletto2016polymer,kang2015confinement}(Fig~\ref{fig:examples}C) as well as protein folding~\cite{grigas2025protein}.

Because rigidity transitions appear to be ubiquitous and important for biological control, it is important to understand how biological systems might regulate properties of their small-scale components to generate large-scale emergent behavior.  Different types of rigidity transitions have different key small-scale control parameters and different responses to perturbations. Therefore, the goal of this article is to discuss the mechanisms and observables associated with rigidity in disordered biological mechanical networks. 

Section~\ref{sec:mechanical_networks} gives detailed examples of biological mechanical networks, and explains how these systems fit within a common mathematical framework. In the next two sections, we describe the fundamentals of rigidity in the simplest limit of zero fluctuations and slow driving. In Section~\ref{sec:constraint_counting} we discuss one type of rigidity transition -- first-order rigidity -- which depends on the connectivity or topology of the network. In Section~\ref{sec:beyond_constraint} we discuss a second type of rigidity transition -- second order rigidity -- which depends on the geometry of the network. In Section~\ref{sec:finite_everything}, we discuss the effects of fluctuations or large deformations, which are common in biological systems, on the mechanical response of systems near a rigidity transition.

\section{Biological structures as mechanical networks}
\label{sec:mechanical_networks}

Biological structures vary widely in composition and mechanical properties. For the remainder of this review, we focus on materials that can be characterized as ``mechanical networks" \ka{modeled as} edges and vertices. \mlm{}

We will first highlight \mlm{a wide range of} examples of biological structures that, perhaps surprisingly, fit well into such a formalism, and then discuss commonalities and differences between them. \mlm{A key point is that that the vertices and edges correspond to different physical structures depending on the type of system. In some cases, the vertex corresponds to the center of mass of an object (e.g. a roughly spherical cell), and the edges are lines between the two cell centers that bisect a point where the two cells come into contact. In other cases, highly deformable cells are approximated as tessellating set of polygons, and so edges are extended cell-cell interfaces, and vertices are points on the polygon. Examples of these different types of representations are highlighted in Fig~\ref{fig:examples}. }

One example is the cytoskeleton, composed of fibers like actin filaments, microtubules, or intermediate filaments, joined by branching proteins or static or dynamic crosslinkers~\cite{head_distinct_2003, fletcher_cell_2010}. To give a sense of the geometry, \emph{in vitro} experiments of model networks of myosin II, actin filaments, and crosslinkers, which mimic physiological responses seen inside cells~\cite{fletcher_cell_2010}, are observed to have a mesh size of about $1 - 3\mu m$ between crosslinks~\cite{mizuno_nonequilibrium_2007}. The typical number of fibers that come together at a vertex -- the coordination number or connectivity --  for cytoskeletal networks is between 3 and 4~\cite{pritchard2014mechanics}, which is important for rigidity considerations as discussed below. To give a sense of force scales, typical stiffnesses for F-actin are about a pN/nm~\cite{pujol2012impact}, and microtubules are similar~\cite{van2009leveraging}. Typical elastic moduli of these networks range from about 0.1-10 kPa~\cite{pritchard2014mechanics}.

\begin{figure}
\includegraphics[scale= 0.8]{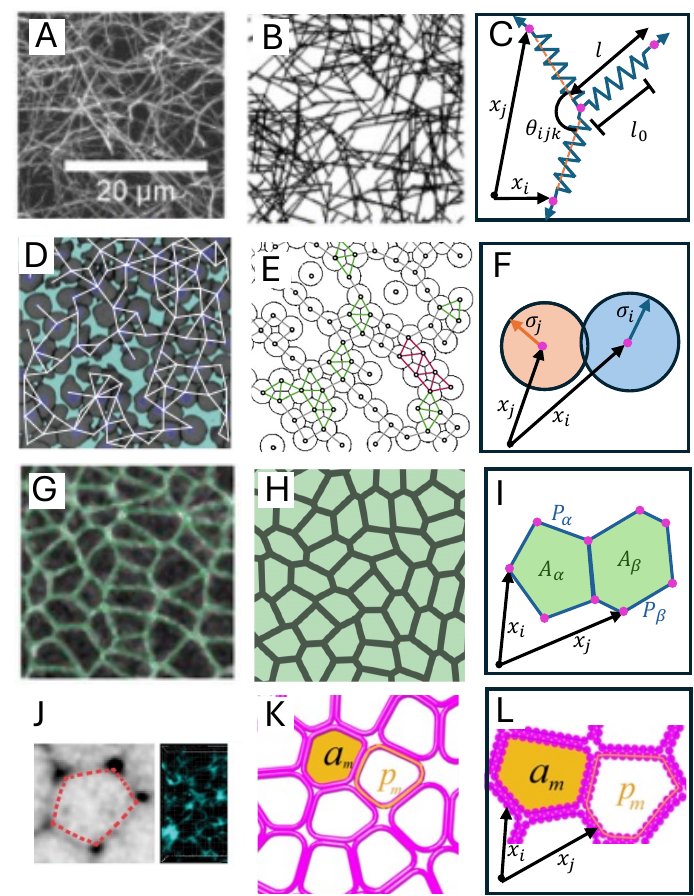}
\caption{{\bf Examples of Biomechanical Networks and Their Corresponding Models} (A) Microscopy images of reconstituted ECM collagen network~\cite{licup_stress_2015}. (B) Mathematical \emph{Mikado} network model of connected springs~\cite{licup_stress_2015}. (C) Fundamental elements of the spring network: degrees of freedom are coordinates $x_i$ of the vertices, constraints are lengths $l$ (and sometimes angles $\theta$) of or between springs, and tunable properties are the rest lengths $l_0$ (and sometimes rest angles $\theta_0$) of or between each spring. (D) Zebrafish blastoderm cell layer during embryo morphogenesis, including overlaid connectivity map~\cite{petridou_rigidity_2021} (E) Simulation snapshot of adhesive spherical cells forming a particle contact network~\cite{petridou_rigidity_2021}. (F) Fundamental elements of a spherical particle network: degrees of freedom are the coordinates $x_i$ of cell centers, constraints are cell-cell contacts, and one of the sets of tunable parameters are the particle radii $\sigma_i$. (G) Microscopy image of confluent \emph{Drosophila} germband tissue during extension~\cite{wang_anisotropy_2020}. (H) Diagram of confluent vertex model, with each cell represented as a discrete polygon. (I) Fundamental elements of a vertex model, where the degrees of freedom are the coordinate locations $x_i$ of the vertices that define polygonal cells, and the constraints are the areas $A_\alpha$ and perimeters $P_\alpha$ of these cells. (J) Confocal sections (inverted) and 3D reconstruction of dextran-labeled extracellular space in the zebrafish tailbud~\cite{mongera_fluid--solid_2018}. (K) Diagram of densely packed but nonconfluent cells as deformable particles~\cite{boromand_jamming_2018}. (L) Fundamental elements of a deformable particle model, where each particle is defined by a ring of tightly-linked vertices that cannot overlap. The degrees of freedom are vertex positions $x_i$ and there are constraints on particle area $a_m$ and perimeter $p_m$~\cite{boromand_jamming_2018}.}
\label{fig:models}
\end{figure}

Another example is the extracellular matrix (ECM), which are networks of fibers outside of cells (Fig~\ref{fig:models}A). ECM has many components, including stiffer proteins like collagen and elastin, as well as softer components comprised of polysaccharides. 
Similar to the cytoskeleton, the fibers are branched and/or connected by static and dynamic crosslinkers; typical mesh sizes between crosslinkers for networks that match physiological properties are about $1-10 \mu m$~\cite{jansen2018role}. The typical connectivity for collagen networks is also between 3 and 4~\cite{jansen2018role, sharma_strain-controlled_2016}. Individual collagen fibers are inferred to have a Young's modulus of about 1 MPa, and the networks have a shear modulus of about 0.1-1 kPa in physiological conditions \cite{jansen2018role}.

        

Both cytoskeletal and ECM networks exhibit interesting rheological responses, including viscoelasticity, mechanical plasticity, and nonlinear elasticity~\cite{head_distinct_2003, chaudhuri_effects_2020}, and both can exhibit significant strain stiffening behavior~\cite{muiznieks_molecular_2013, sharma_strain-controlled_2016} discussed later in this review.

These networks are also well-described by similar mathematical models: spring networks with bending energies (Fig~\ref{fig:models}B,C):
\begin{align}
    E = \frac{K_L}{2}\sum_{\langle i,j \rangle}(l_{ij} - l_0)^2 + \frac{K_\theta}{2}\sum_{\langle i,j,k \rangle}(\theta_{ijk} - \theta_0)^2 \label{spring network bending energy}
\end{align}
where the actual lengths $l_{ij} = \sqrt{ \sum_{\mu} \left( x_{i\mu} - x_{j\mu} \right)^2}$ are the distance between vertices $i$ and $j$,  $x_{i \mu}$ represents the Cartesian coordinates of vertex $i$, with $\mu \in \{x,y,z\}$. Here $l_0$ is the rest length of the spring, $K_L$ is the stiffness of the linear springs, and $K_{\theta}$ is the stiffness of the angular springs with actual angles $\theta_{ijk}$ and rest angle $\theta_0$ \cite{broedersz_criticality_2011}. 

In many systems, the bending stiffness is much smaller than the stretching stiffness, and in that case there is a crossover from a very soft, bending-dominated regime at low strains to a very stiff, stretching-dominated regime at high strains~\cite{broedersz_criticality_2011}. To predict this behavior, it is often sufficient to approximate the network as a central force network of linear springs and neglect the bending terms in Eq.~\ref{spring network bending energy}~\cite{sharma_strain-controlled_2016}.

Groups of cells can also be thought of as mechanical networks, with edges representing either effective interactions between cell centers, or cell-cell contacts. Edge lengths are about $10 \mu m$ -- the length scale of a typical animal cell, with typical elastic moduli measurements on the order of 0.1-0.8 kPa~\cite{angelini_glass-like_2011, serwane_vivo_2017}. These networks also exhibit non-trivial rheological behavior, including glassy dynamics~\cite{Angelini} and complex viscoelastic responses~\cite{serwane_vivo_2017}.

For loose packings of cells with rounded, uniform shapes (Fig~\ref{fig:models}(D)), particle-based descriptions are useful, i.e. isotropic two-body interactions with a finite cutoff, or slightly squishy, possibly adhesive spherical particles. 

For example, interactions between pairs of frictionless foam bubbles are well-represented by a one-sided harmonic potential:
\begin{equation}
    E^{tot} =  
    \begin{cases}
        \frac{1}{2}\sum_{<ij>} k (1 - \frac{l_{ij}}{\sigma_{i}+\sigma_{j}})^2, \quad  l_{ij} <  (\sigma_i + \sigma_j), \\
     0 \quad \quad \textit{otherwise},
    \end{cases}
    \label{eq:particlepotential}
\end{equation}
 where $l_{ij}$ is the distance between particle centers at $x_{i \mu}$ and $x_{j \nu}$. The particle stiffness is $k$ and the particle radius is $\sigma_{i}$~\cite{durian1995foam}. In this case, the mechanical network is the network of bonds created by cell-cell contacts (Fig~\ref{fig:models}E,F).
 
 Similar interaction potentials, sometimes with spring backbones, are also used to model crowded cytosol~\cite{mcguffee2010diffusion}, chromatin~\cite{michieletto2016polymer}, and even some protein condensates~\cite{benayad2020simulation}. 
One downside to this modeling approach is that the potential does not depend on how much the particles interact or deform, which becomes a poor approximation at higher densities.

To address this limitation, another class of \emph{vertex} models has been developed to describe confluent tissues -- where cells are so dense that there are no gaps or overlaps between cells -- which is common in epithelial cell types (Fig~\ref{fig:models}G).

Vertex models define tissues as a network of edges connected at (typically three-fold coordinated) vertices to form a tessellation of polygons~\cite{nagai_dynamic_2001}.  The degrees of freedom for this network are the positions of the vertices (Fig~\ref{fig:models}H,I). In its standard 2D form~\cite{Farhadifar}, the mechanical energy is a function of the cell's cross-sectional area $A_{\alpha}$ and its cross-sectional perimeter $P_{\alpha}$:
\begin{equation}{\label{vm}}
    E = \sum_\alpha \left[ K_A(A_\alpha - A_0)^2 + K_P(P_\alpha -P_0)^2\right].
\end{equation}
where $A_0$ is a \emph{target area} that the cell maintains via regulation of its volume and height fluctuations in a monolayer, and $P_0$ is the \emph{target perimeter} that the cell maintains via regulation of adhesion molecules and an active actin/myosin gel at the cell cortex. $K_A$ and $K_P$ are moduli that govern how tightly the target area and perimeter are regulated. Vertex models in three dimensions are similar; cells are represented as polyhedrons that fill space, and cell-cell contacts are the polygon faces. The degrees of freedom are still the vertices, but the constraints are now proportional to cell surface area and volume~\cite{honda_three-dimensional_2004, merkel_geometrically_2018}. 



Another variation is an active tension model~\cite{noll_active_2017}, where each edge has its own line tension, and the area is constrained:
\begin{equation}{\label{atn}}
    E = \sum_\alpha K_A(A_\alpha - A_0)^2 + \sum_{<i,j>} \Lambda l_{ij}.
\end{equation}
In some version of these models, $K_A \to \infty$ so the area is always equal to its target value.


A limitation of vertex models is that it is not possible for the edges to slide past one another or for gaps to open up between cells. To remedy this, researchers have introduced deformable particle, active foam, and phase-field models that account for changing cell shapes in systems where cells are more tightly packed than in particle-based models, and yet gaps still remain between cells (Fig~\ref{fig:models}J). 2D deformable particle models approximate a single cell as a ring of vertices connected by edges~\cite{boromand_jamming_2018}, with quadratic penalties for changing the perimeter and area of each cell, similar to Eq~\ref{vm} . In addition, there are strong penalties for nearly overlapping vertices that generate an effective repulsion between cell-cell interfaces (Fig~\ref{fig:models}K,L).  Active foam models represent cells as a series of networks composed of vertices connected by straight edges or arcs that form minimal surfaces, with an energy functional similar to Eq.~\ref{atn}; cell-cell contacts possess a distinct surface tension from cell-gap interfaces~\cite{ActiveFoam}. Phase-field models are similar in structure to active foams, and represent the network of cell interfaces by level sets of a continuous function~\cite{nonomura2012study,loewe_solid-liquid_2020}. 

Although biomechanical network models are quite different in how they represent the energy and degrees of freedom, there is an overarching similarity. They all have two types of variables: the vertices, which are the physical degrees of freedom in the mathematical model, and the constraints that restrict those degrees of freedom (Fig~\ref{fig:models} right column).  In general, the mechanical energy is expressed as a function of the constraints. Such a description is also used in many other mechanical systems (e.g. bar-joint networks, tensegrity models, origami). 
To find mechanically stable states, one evolves the vertices to reduce the energy until a local minimum is found. Then the emergent mechanical properties of the system can be obtained by studying perturbations away from these mechanically stable states.

\mlm{Another similarity is that in all of these networks, we can define a number density of vertices per unit area (2D) or volume (3D), e.g. $N_v/A_{box}$, which has units of inverse length squared in 2D and is discussed in Section~\ref{sec:beyond_constraint}(C) below. In some cases, especially in the context of the jamming transition discussed in Section~\ref{sec:constraint_counting}, it is useful to define a dimensionless quantity, called the packing fraction, which for soft spheres is defined as the space taken up by the particles divided by the space in the box, e.g. $\phi = N_v \pi r^2/A_{box}$ in 2D. Notice that for soft spheres the packing fraction scales linearly with the number density of vertices, but this will not generally be true for other types of mechanical networks.} 

\section{First order rigidity transitions}
\label{sec:constraint_counting}
In recent work, Petridou and collaborators discovered that the tissue that comprises the early zebrafish embryo undergoes a solid-to-fluid transition at a precise point in development -- when the tissue undergoes a very large deformation to set up the anterior-posterior axis. This change in material properties is associated with a change in the network of contacts between cells. In the solid phase, the tissue has a percolating network of cell-cell contacts, and then a decrease in cell-cell adhesion opens up more gaps between the cells, slightly changing the average number of contacts and dramatically destroying the percolating contact network, resulting a fluid-like material that flows easily. A key feature is that the components make contacts only within a finite interaction range, and break contacts outside that range. An example of such a potential is given by Eq.~\ref{eq:particlepotential}. 

The mathematical framework that describes this behavior is that of first-order rigidity transitions. \mlm{Roughly speaking, a first-order rigidity transition occurs when the number of degrees of freedom equals the number of constraints, which is why it is also known as \emph{constraint counting}; it is the simplest and perhaps most common criteria for determining whether a material is rigid. More technically, this criterion considers whether first-order perturbations to the constraints cost energy, which ultimately leads to equations that count the rows and columns of a matrix of first derivatives. However, as we discuss in Section~\ref{sec:beyond_constraint}, a material can be rigid even if this first-order check fails.} 

To formalize our discussion of constraints at the end of the last section, for each contact or bond $\alpha$, we define a constraint $f_\alpha = \sqrt{k_\alpha} \left( 1 - \frac{l_{ij}}{\sigma_{i}+\sigma_{j}}\right)$ and we see that the energy functional can be written as a sum over the bonds or quadratic constraints,
\begin{equation}
    E = \frac{1}{2}\sum_{\alpha} f_\alpha^2, \label{EnergyConstraints}
\end{equation}
which is also true for many other systems of interest, including bar-joint networks for engineered structures.

Our goal is to predict whether a network of constrained vertices is rigid or not. In general, this is a difficult (NP-hard~\cite{abbott_generalizations_2008}) problem. In some cases, it is possible to use simpler arguments to prove a structure is rigid. These simpler conditions for rigidity are sufficient, but not necessary, to prove rigidity, resulting in a Venn diagram of increasingly complex conditions for rigidity (Fig~\ref{fig:constraints}A).

In this section, we review the simplest (and most restrictive) condition: first-order rigidity, defined using the \emph{rigidity matrix} $R$. $R$ is an $N_b$ by $N_v d$ matrix that describes how infinitesimal changes to the vertices (i.e the $N_v d$ degrees of freedom) alter the edge lengths (i.e, the $N_b$ bonds or constraints): $R_{\alpha i\mu} = \partial f_\alpha / \partial x_{i\mu}$, where $i=1,...,N_v$ indexes the vertices and $\mu = 1,..,d$ indexes the spatial dimensions. It therefore describes resulting first-order change to the constraints:
\begin{align}
    \delta f_\alpha = \sum_{i\mu} \frac{\partial f_\alpha}{\partial x_{i\mu}}\delta x_{i\mu} = \sum_{i\mu} R_{\alpha i\mu} \delta x_{i\mu} = 0, \label{FirstOrder}
\end{align}

Modes in null($R$) are displacements of the positions that do not impact the constraints (and therefore the energy) to linear order, and are called \emph{linear zero modes} (LZMs, physics literature~\cite{damavandi_energetic_2021,lubensky_phonons_2015}), or first order flexes (engineering, applied math literature~\cite{connelly_rigidity_1980, connelly_higher-order_1994, holmes-cerfon_almost-rigidity_2021}).  

$R$ also relates the tensions on the edges of the network to the resulting forces on the vertices $F_{i\mu}$:
\begin{align}
    -F_{i\mu} = \frac{\partial E}{\partial x_{i\mu}} = \sum_\alpha \frac{\partial E}{\partial f_\alpha}\frac{\partial f_\alpha}{\partial x_{i\mu}} = \sum_\alpha f_\alpha R_{\alpha i\mu} . \label{Equilibrium}
\end{align}

\mlm{Although this derivation requires that the energy can be written as a sum over constraints, as in Eq.~\ref{EnergyConstraints}, Section~\ref{sec:mechanical_networks} highlights that it is a reasonable starting assumption for many biological structures.}
If the constraints are not all satisfied $(f_\alpha \neq 0)$, we say the network is prestressed, and Eq. (\ref{Equilibrium}) implies that $f_\alpha$ must be in the left nullspace of $R$ in order for the network to be in equilibrium. 
Modes in the left nullspace of $R$, or null($R^{T}$), are stresses on the bonds that don't cause any displacements to first order and are called states of self stress, denoted $\sigma$.

To see this directly from the definition of the rigidity matrix, one can write the rank-nullity theorem for $R$ and $R^{T}$. $N_v d$ = rank($R$) + null($R$), and $N_b =$ rank($R^{T}$) + null($R^{T}$). Let's write the total number of bonds in the system $N_b = zN_v/2$, where $z$ is the average number of contacts per particle, $N_v$ is the number of particles, and we divide by $2$ because each bond is shared by two particles. Since rank($R$) =rank($R^{T}$):
\begin{equation}
    N_v d - N_0 = zN_v/2 - N_{ss},
    \label{eq:constraint_counting}
\end{equation}
where $N_0$ is the number of linear zero modes and $N_{ss}$ is the number of states of self-stress. This is called Maxwell-Calladine constraint counting~\cite{maxwell_l_1864, calladine_buckminster_1978}. This mathematical framework leads us to expect that materials will be rigid when the number of constraints is equal to or greater than the number of degrees of freedom. In that case, changing the vertices necessarily changes the edge lengths, which costs energy. This is also called ``first-order rigidity" because Eq~\ref{FirstOrder} expands only to first order perturbations in the constraints.  

This framework correctly predicts the mechanics of disordered jammed repulsive spheres: at the jamming transition there are no linear zero modes and only one state of self stress, so that for large $N_v$, $z \sim 2 d$, which is called the isostatic condition.  In an experiment or simulation, $z$ is difficult to access directly; instead one usually controls the pressure or boundary conditions, which in turn govern the packing fraction $\phi$. Analogous to the 2D quantity discussed in Section~\ref{sec:mechanical_networks}, in 3D $\phi$ is the ratio of volume taken up by the particles relative to the total volume of the box. At densities below a critical packing fraction ($\phi_J \sim 0.64$ in 3D for frictionless spheres), particles can rearrange so as not to touch and $z=0$~\cite{o2003jamming}. At the critical packing fraction $\phi_J$, the coordination number jumps to exactly $z=2d$, and then $z$ continues to increase with density beyond $\phi_J$, controlling scaling properties of the elastic moduli~\cite{goodrich2016scaling}.

But exactly how is the emergent property $z$ related to $\phi$? This turns out to be an incredibly rich and difficult question. For hard spheres in infinite dimensions (instead of the usual 2 or 3), this question can be solved exactly using a replica symmetry breaking approach~\cite{parisi2010mean}. Useful concepts that arise from that theory include the fact that jammed systems are in a marginally stable~\cite{muller2015marginal} Gardner phase~\cite{charbonneau2017glass}, resulting in avalanche dynamics under finite applied forces, discussed in Section~\ref{sec:finite_everything}. Somewhat surprisingly, some exact predictions for scaling exponents from the infinite-dimensional theory survive to $d=2$ and $d=3$~\cite{charbonneau2017glass}. On the other hand, the infinite-dimensional theory does not correctly predict other features of low-dimensional jammed solids, such as the density of vibrational states that control plastic flow~\cite{berthier2025yielding}. 

In addition to these theoretical results, there is an extensive body of numerical work in two and three dimensions, with many different types of interaction potentials, characterizing power law scaling relations between $z$ and $\phi$~\cite{o2003jamming}. Ultimately, this confirms that in sphere packings, one can control the transition by controlling the packing fraction, which in turn controls $z$.

Although repulsive spheres packing are typically tuned by changing boundary conditions or applied pressure~\cite{o2003jamming}, the degree of adhesion in adhesive packings can control the packing fraction network connectivity~\cite{koeze2018sticky, grigas2025protein}. In the work by Petridou et al, the packing density of sphere-like cells is controlled by effective adhesion between the cells; as the adhesion is decreased, the cells pack slightly less tightly together, reducing the packing fraction and crossing the jamming transition, transitioning from $z \sim z_c = N_vd$ in the solid phase to almost no contacts in the fluid phase.

    


 
\section{Second order rigidity transitions}
\label{sec:beyond_constraint}
\subsection{Biological examples where first-order rigidity fails}

Many biomechanical networks are ``undercoordinated", meaning they do not have enough constraints to satisfy constraint counting requirements for first-order rigidity. Despite this, in some parameter ranges or under certain boundary conditions, they are observed to be rigid.


    
For example, 3D collagen networks are observed to have on average between three and four fiber branches at each vertex~\cite{sharma_strain-controlled_2016}, which is significantly less than the isostatic coordination of $2d = 6$ in three dimensions.  However, these networks switch from being quite floppy to stiffening by orders of magnitude under an applied shear or dilation strain~\cite{licup_stress_2015, sharma_strain-controlled_2016}. Moreover, their mechanics displays a lack of sensitivity to the concentration of collagen fibers~\cite{licup_stress_2015}, suggesting that network connectivity is not dominating their mechanical behavior. Mathematical models for fiber networks (e.g. Eq~\ref{spring network bending energy}) recapitulate this behavior~\cite{mackintosh_elasticity_1995,sharma_strain-controlled_2016}.


Similarly, experiments have demonstrated that 2D epithelial layers can undergo a fluid-solid transition, including cultured human bronchial epithelial tissue~\cite{park_unjamming_2015}, cultured MDCK monolayers~\cite{devany_cell_2021}, and \emph{in vivo} Drosophila germband during convergent extension~\cite{wang_anisotropy_2020}, and that these transitions are governed by cell shapes as predicted by vertex models. 

However, vertex models are typically also underconstrained. \mlm{For completeness, the remainder of this paragraph briefly demonstrates this assertion using standard results from the field of topology; it is not necessary to follow the remainder of this review.}
The 2D vertex model has two constraints per cell (one on the area, one on the perimeter). In flat 2D cellular systems, topological constraints ensure that the average number of neighbors per cell is six ($n=6$), and the Euler equation is satisfied: $0 = V - E + N$, where $V, E, N$ are the total number of vertices, edges, and cells in the tiling~\cite{Weaire}. Since each edge is shared by two cells, $E = Nn/2$, the average number of unique edges per cell is three, and $V = N(n/2 -1)$, so the number of unique vertices per cell is two (i.e., a two-atom basis generates a honeycomb). Because each of those vertices can move in either the $x-$ or $y-$ direction, there are $4$ degrees of freedom per cell, compared to two constraints per cell; i.e., vertex models are also generally underconstrained.

Nevertheless, vertex models are observed to transition from floppy to rigid based on their geometry.  This transition can be controlled by an internal parameter, the target shape index ($p_0 = \frac{P_0}{\sqrt{A_0}}$)~\cite{bi_density-independent_2015}, or an external parameter, the dilational or shear strain~\cite{merkel_minimal-length_2019}.   For a fixed box size, vertex model networks are rigid below a characteristic value of $p_0^* \simeq 3.81$; above it, they are floppy~\cite{Bi2015}.  See Ref~\cite{merkel_minimal-length_2019} for a discussion of constraint counting in other vertex-like models, and Ref~\cite{damavandi_universal} for a discussion of rigidity transitions in a wide range of vertex models.

\begin{figure}
\centering
\includegraphics[scale=0.55]{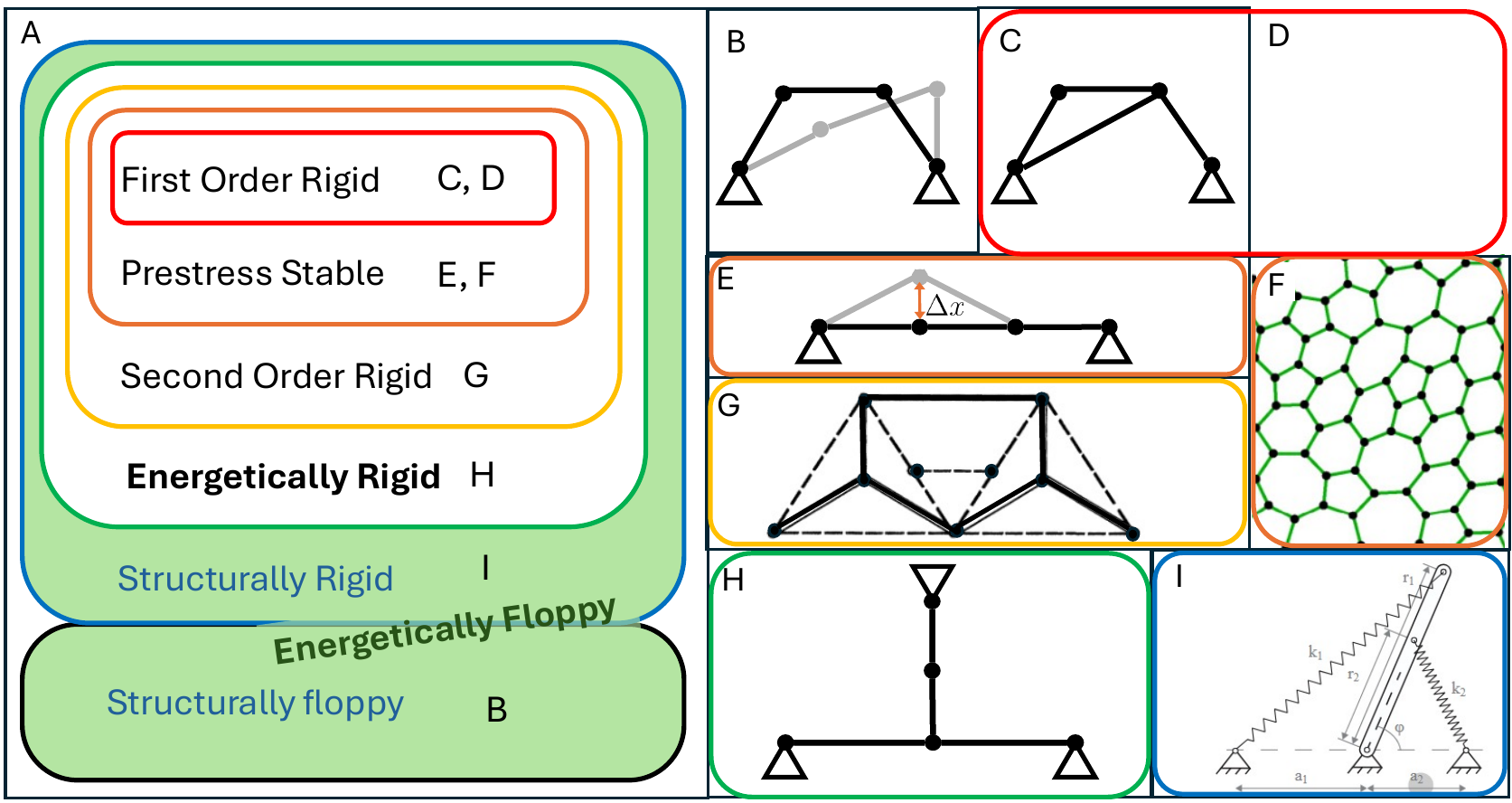}
\caption{{\bf Examples of different conditions for rigidity} (A) A Venn diagram illustrating the hierarchy of the various definitions of rigidity, e.g. all First Order Rigid systems are pre-stress stable, but not vice-versa. (B-J) Examples of systems with different types of rigidity. Dots represent vertices, and triangles represent hinges that can rotate but not translate. Solid lines represent bars that resist compression or extensions. Dashed lines represent cables that cannot resist compression. (B) A generic three-bar linkage, which is both structurally and energetically floppy, with the floppy mode shown in gray. (C) Adding another edge to the three-bar linkage makes it first order rigid. (D) A sphere packing above the isostatic point is first order rigid. (E) A three-bar linkage in a critical configuration with a state of self-stress which makes it prestress stable but not first order rigid. (F) A disordered 3-coordinated spring network in a configuration with a state of self-stress is prestress stable but not first order rigid. (G) A bar and cable framework which is second order rigid but not prestress stable~\cite{connelly_second-order_1996}. (H) A linkage which is energetically rigid at 4th order, but is not second order rigid. (I) A framework which is structurally rigid, but is energetically floppy when the spring constants are precisely tuned~\cite{schenk_zero_2014}.}
\label{fig:constraints}
\end{figure}

        
\tah{\subsection{Beyond Constraint Counting}}
These examples confirm that first-order rigid networks are only a subset of all rigid structures. 
 Maxwell's original work on the stiffness of frameworks \cite{maxwell_l_1864}, demonstrated that there could be special configurations of underconstrained systems that are still rigid, in which ``certain conditions must be fulfilled, rendering the case one of a maximum or minimum value of one or more of its lines." 
These special frameworks include Buckminster Fuller's \emph{tensegrity} structures, which are constructed from rigid struts and tensioned cables (Fig~\ref{fig:constraints}G). 
Perhaps the simplest example is the three-bar linkage \tah{(Fig~\ref{fig:constraints}B) with four degrees of freedom (two per vertex) and three constraints.}. In a generic configuration (Fig~\ref{fig:constraints}B), the linkage has no self-stress \tah{because any assignment of tension to the bars would lead to non-zero forces on the vertices. Constraint counting (Eq. \ref{eq:constraint_counting}) implies that there is a single LZM, since
\begin{align}
N_0 = N_{\text{dof}} - N_{\text{const}} + N_{ss} = 4-3+0=1.
\end{align}
This LZM corresponds with the shearing motion shown in gray in Fig~\ref{fig:constraints}B which costs zero energy, implying the framework is floppy. There are two ways for the three-bar linkage to become rigid. If an additional bar is included as in Fig~\ref{fig:constraints}C, the framework still does not possess a self-stress, so constraint counting now implies that there are no LZM's and the framework is first-order rigid. Alternatively, without adding any new constraints one could shorten the existing bars until they are colinear as in Fig~\ref{fig:constraints}E. In this configuration the framework now possesses a self-stress where all three bars are under equal tension, so constraint counting implies there are two LZM's:
\begin{align}
N_0 = N_{\text{dof}} - N_{\text{const}} + N_{ss} = 4-3+1=2,
\end{align}
which correspond to vertical motions of the two interior vertices (one of which is shown in gray in Fig~\ref{fig:constraints}E).} Because these motions are orthogonal to the horizontal edges, they change the lengths of the edges only to second order. In other words, if a previously horizontal spring of length $l$ has one its vertices displaced vertically by a distance $\Delta x$ \tah{(indicated in Fig~\ref{fig:constraints}E)}, the new length of a spring $l_{new}$ is the hypotenuse of a right triangle with sides $\Delta x$ and $l$:
\begin{equation}
l_{new} = \sqrt{l^2 + \Delta x^2} \simeq l + \frac{\Delta x^2}{l}.
\end{equation}
I.e., the change in length is proportional to $\Delta x^2$, and therefore the first-order equation, Eq~\ref{FirstOrder}, does not account for it. Thus, while the collinear configuration has more linear zero modes than a generic configuration, these modes are stabilized by the state of self-stress which makes the linkage rigid at second order.

This idea was rigorously formalized with the definition of second order rigidity~\cite{connelly_rigidity_1980, connelly_second-order_1996}. \tah{While the three-bar linkage provides a sufficient intuitive explanation of this phenomenon, in the following subsection} we review this formal derivation in some detail \tah{for the interested reader}, as second-order rigidity occurs regularly in many biomechanical networks.


\subsubsection{Mathematical formalism for second order rigidity}

 \tah{In the following derivations w}e use a more general notation from~\cite{damavandi_energetic_2021}, as~\cite{connelly_rigidity_1980, connelly_second-order_1996} were written specifically for the context of bar-joint frameworks. As before, we assume a set of degrees of freedom $\{x_i\}$ and a set of constraints $\{f_\alpha\}$. Given an analytic deformation to the degrees of freedom $x_i(t)$, we can calculate the change to the constraints up to second order in the parameter $t$:
\begin{equation}
    \delta f_\alpha = \left[\sum_i \frac{\partial f_\alpha}{\partial x_i}\dot{x}_i \right]t +\frac{1}{2}\left[\sum_i \frac{\partial f_\alpha}{\partial x_i}\ddot{x}_i + \sum_{ij} \frac{\partial^2 f_\alpha}{\partial x_i \partial x_j}\dot{x}_i \dot{x}_j\right]t^2 + \mathcal{O}(t^3). \label{second order constraint expansion}
\end{equation}
To second order we can describe the deformation by the first two terms in its Taylor series:
\begin{align}
    x_i(t) = x_i(0) + \dot{x}_i t + \frac{1}{2}\ddot{x}_i t^2 + \mathcal{O}(t^3).
\end{align}
We say that the ordered pair $(\dot{x}_i, \ddot{x}_i)$ is a second order flex if it does not change the constraints at second order; that is, if it satisfies
\begin{align}
    \sum_i \frac{\partial f_\alpha}{\partial x_i}\dot{x}_i &= 0  \label{second order rigidity condition 1} \\
    \sum_i \frac{\partial f_\alpha}{\partial x_i}\ddot{x}_i + \sum_{ij} \frac{\partial^2 f_\alpha}{\partial x_i \partial x_j}\dot{x}_i \dot{x}_j &= 0. \label{second order rigidity condition 2}
\end{align}
A system is said to be second order rigid if every second order flex has a trivial linear component; that is, $\dot{x}_i$ is either zero or a rigid body motion. Ref~\cite{connelly_second-order_1996} proved that this condition is sufficient to guarantee structural rigidity to all orders.

How can we tell when a given system has a nontrivial second order flex $(\dot{x}_i, \ddot{x}_i)$? Eq~\ref{second order rigidity condition 1} implies that $\dot{x}_i$ from this pair is a nontrivial linear zero mode.  Must the pair have a state of self stress? Consider an underconstrained system that does not possess a state of self-stress\tah{: in this case the rigidity matrix $\partial f_\alpha / \partial x_i$ is full rank. Then for any choice of $\dot{x}_i$ Eq \ref{second order rigidity condition 2} is a linear equation in $\ddot{x}_i$ of the form $A\vec{x} = \vec{b}$, which is guaranteed to have a solution if $A$ is full rank. Thus if the system does not have a self-stress, for any LZM $\dot{x}_i$} there is always a choice for the second order term $\ddot{x}_i$ that will satisfy Eq \ref{second order rigidity condition 2} to produce a nontrivial second order flex. Thus, such a system cannot be second order rigid. 


Therefore, we further consider only flexes that possesses $N_{ss} > 0$ independent states of self-stress. If there are $m$ total constraints, we can choose an orthonormal basis $\{ \sigma_\alpha^1, ... , \sigma_\alpha^{N_{ss}}, e_\alpha^1, ..., e_\alpha^{m-N_{ss}} \}$ for $\mathcal{R}^m$, where the $\sigma$'s span the space of states of self-stresses 
\begin{align}
\text{span}(\sigma_\alpha^1, ... , \sigma_\alpha^{N_{ss}}) = \text{null}(R^T), 
\end{align}
and the $e$'s span the rest of $\mathcal{R}^m$. Eq \ref{second order rigidity condition 1} still implies that $\dot{x}_i$ is a nontrivial LZM, and we can project Eq \ref{second order rigidity condition 2} onto our orthonormal 
basis, which after some algebra results~\cite{damavandi_energetic_2021} in a test for second-order rigidity:
%
a system is second order rigid if there is no nontrivial linear zero mode $\dot{x}_i$ that satisfies
\begin{align}
    \sum_{ij\alpha} \sigma_{\alpha} \frac{\partial^2 f_\alpha}{\partial x_i \partial x_j}\dot{x}_i \dot{x}_j = 0, \label{second order rigidity stress test}
\end{align}
for all states of self-stress $\sigma_\alpha$. This definition can be used to prove that underconstrained spring networks and vertex models are second-order rigid~\cite{damavandi_energetic_2022-1}.

While second order rigidity is sufficient to imply structural rigidity, it is not necessary. Fig~\ref{fig:constraints}(H) shows a linkage that is not second order rigid, but has a second order flex that changes the edge lengths at 4th order. One might think that the above formalism could be extended to obtain higher-order rigidity criteria, but it is not straighforward\cite{connelly_higher-order_1994}. Very recent work~\cite{gortler2025higher} has developed an energy-based formalism to handle these cases.

Eq \ref{second order rigidity stress test} is generally a difficult condition to test. However, it often suffices to check a weaker condition, \emph{prestress stability}~\cite{connelly_second-order_1996}. If there is a particular state of self-stress $\sigma_\alpha$ such that the associated ``prestress matrix"
\begin{align}
    P_{ij} = \sum_{\alpha} \sigma_{\alpha} \frac{\partial^2 f_\alpha}{\partial x_i \partial x_j}, \label{prestress matrix definition}
\end{align}
is positive-definite on the subspace of nontrivial linear zero modes, then the system is said to be prestress stable. Prestress stability implies second order rigidity. For any potential second order flex to satisfy Eq \ref{second order rigidity stress test} requires $\sum_{ij}P_{ij}\dot{x}_i \dot{x}_j = 0$, which is impossible if the prestress matrix is positive definite on the nontrivial LZM's. However, the converse is not true, as there are frameworks that are second order rigid but not prestress stable (Fig~\ref{fig:constraints}(G)). If a system has only a single independent state of self-stress, then prestress stability is equivalent to second order rigidity.

All of these definitions of rigidity are focused on the structural rigidity of a system: whether or not there are any deformations that leave the individual constraints unchanged. However, physicists usually define rigidity in terms of the energetics of a system \cite{damavandi_energetic_2021} -- a material is floppy if it costs zero energy to deform it. This is usually equivalent to asking whether the Hessian matrix -- the derivatives of the energy with respect to the vertices $H_{i \mu j \nu} = \partial E / (\partial x_{i \mu} \partial x_{j \nu}$) -- is positive definite; i.e., all the eigenvalues of the matrix are strictly greater than zero. However, as we will see below there are special cases where the Hessian has eigenvalues that are zero but the corresponding eigenvectors cost energy at higher order~\cite{damavandi_energetic_2021}.

If there is an eigenvector of the Hessian matrix that has a zero eigenvalue and that eigenvector overlaps with a global shear, then the shear modulus $G$ is zero. Otherwise, the shear modulus $G$ is simply a weighted sum over the eigenvectors of the Hessian matrix~\cite{merkel_geometrically_2018}, and so the shear modulus is generically finite and positive when $H_{i \mu j \nu}$ is positive definite. This is why physicists use finite $G$ as a proxy for defining rigidity.

If a system has a quadratic energy cost associated with each constraint (as in Eq \ref{EnergyConstraints}), then the Hessian matrix can be written as:
\begin{align}
    \frac{\partial^2 E}{\partial x_i \partial x_j} &= \sum_\alpha \frac{\partial f_\alpha}{\partial x_i} \frac{\partial f_\alpha}{\partial x_j} + \sum_{\alpha} f_{\alpha} \frac{\partial^2 f_\alpha}{\partial x_i \partial x_j} \\
    &= (R^T R)_{ij} + P_{ij}. \label{Hessian}
\end{align}
The Hessian matrix is made up of two terms: a Gram term (or stiffness matrix~\cite{connelly_second-order_1996}) and the prestress term. Because the Gram term is the square of the rigidity matrix, it is positive semi-definite with a nullspace spanned by the linear zero modes. If the system is prestressed ($f_\alpha \neq 0$), mechanical equilibrium implies that the system must possess at least one independent state of self-stress $\sigma_\alpha$ such that $f_\alpha = C\sigma_\alpha$ for some $C > 0$. Then the lowest order energy cost for a linear zero mode $\dot{x}_i$ is given by
\begin{align}
    \delta E = \sum_{ij} \dot{x}_i P_{ij} \dot{x}_j =  C \sum_{ij\alpha} \sigma_{\alpha} \frac{\partial^2 f_\alpha}{\partial x_i \partial x_j}\dot{x}_i \dot{x}_j.  \label{LZM energy cost}
\end{align}
In order for the system to be in a stable equilibrium, this energy cost must be positive for all nontrivial LZM's, which implies that the system is prestress stable and hence rigid. If the system is not prestressed ($f_\alpha = 0$), then the prestress term vanishes. In this case, all linear zero modes will also be zero modes of the Hessian, so many typical measures of rigidity will vanish (e.g. $G = 0$). However, such systems could still be rigid. One can imagine starting with a prestressed rigid system and scaling the magnitude of prestresses to zero while keeping the configuration fixed (setting $C=0$). The prestress term vanishes so the Hessian still possesses many zero modes, but the existence of states of self-stress produces an energy cost at 4th order:
\begin{align}
    \delta E \propto \sum_{I=1}^{N_{ss}} \left[ \sum_{ij\alpha} \sigma^I_{\alpha} \frac{\partial^2 f_\alpha}{\partial x_i \partial x_j}\dot{x}_i \dot{x}_j \right]^2.  \label{quadratic energy cost}
\end{align}
This is an example of one of the special cases where $G=0$ even though the system is energetically rigid, and it occurs regularly in underconstrained biomechanical networks right at the onset of the rigidity transition.

In addition, it is possible to design systems that are energetically floppy even if they are structurally rigid -- these must have multiple states of self stress that interact in a non-trival way so that the energy remains constant even when different bonds contract or extend, e.g. Fig~\ref{fig:constraints}(J)~\cite{schenk_zero_2014}. Fig~\ref{fig:constraints}(A) summarizes this hierarchy of different tests and types of rigidity. Jammed packings are first-order rigid; vertex models and underconstrained fiber networks are second-order rigid, and there are some subtle difference between other definitions of rigidity that have been exploited in man-made structures. An interesting open question is whether any of these more exotic structures are exploited in biology.


\subsection{Why are so many underconstrained biomechanical networks rigid?}

Harkening back to the Maxwell quote that introduced this section, although it is possible for underconstrained systems to become rigid, one expects such states to be special or non-generic. So why do rigid states appear to show up so often in underconstrained biomechanical networks?
A useful observation is that all of our second-order rigid biomechanical networks can exhibit a geometric incompatibility~\cite{moshe2018,vermeulen_geometry_2017, merkel_minimal-length_2019}.  That is, there is one energetic length scale set by the constraints (e.g. the rest lengths of springs $l_0$, target cell perimeters $p_0$, etc), and a second intrinsic length scale set by the number density of vertices in the box (e.g. $l_N = \sqrt{A_{box}/N_v}$ in 2D). In disordered networks, if the intrinsic lengthscale $l_N$ becomes larger than the energetic lengthscale (e.g. $p_0/n$ with $n$ the number of edges in a polygon), the system can no longer satisfy the constraints, the energy becomes finite, and the system becomes rigid.  This geometric incompatibility can be induced by extending boundaries of the box under fixed internal constraints (e.g. dialational strain that increases $A_{box}$ at fixed $N_v$) or instead by fixing the boundaries and shrinking the energetic lengthscale (e.g. reducing the target shape index $p_0$ in vertex models).


On the floppy side of this transition these systems possess no states of self-stress and an extensive number of zero modes. At a critical value of ratio between the intrinsic and energetic lengthscales, they gain a state of self-stress and become second order rigid, but are not yet prestressed. Because the prestress matrix is zero, the Hessian still has zero modes and the shear moduli vanish (in the thermodynamic limit), but these modes cost energy at 4th order. This critical point corresponds to a choice of geometry that is ``just barely" compatible with the system's boundary conditions. 

On the rigid side of the transition the preferred geometry is impossible to achieve, which leads to finite amounts of prestress. Now the system is prestress stable and the Hessian is positive definite (aside from trivial rigid body rotations and translations). Because the linear zero modes are stabilized by the prestress, many mechanical properties scale with the prestress and hence with the preferred geometry near this critical point, including the low frequency portion of the density of states, which also controls the elastic moduli \cite{merkel_minimal-length_2019,damavandi_energetic_2022-1}.

In the case of central force networks, one can make this statement precise by demonstrating that the set of critically rigid configurations is a manifold almost everywhere, and that the manifold is co-dimension one in the configuration space -- specifically, the space of the squared edge lengths of the network~\cite{hain_optimizing_2025}. 
Because the manifold is co-dimension one, changing the internal parameters or the external boundary conditions along any generic trajectory will eventually intersect the manifold, signifying a second order rigidity transition. 

\tah{This is a key mathematical result that explains why so many underconstrained biomechanical networks are observed to be rigid. They are able to tune their geometric incompatibility -- either via local strains or adjusting the energetic constraint lengthscale -- and such tuning will \textbf{generically} cross the critical rigidity manifold because it is co-dimension one. This also gives rise to universal features in the geometry and rheology around the critical point, regardless of the model details~\cite{damavandi_universal_2022}. In the following section we review some of the mathematical details of this critical manifold in central force networks for the interested reader.}

\subsubsection{Mathematical Details of the Critical Manifold in Central Force Networks}

\tah{Central force networks are a convenient model in which to study the critical manifold, as it} can be \tah{directly} parameterized with the state of self-stress that appears at the critical point, though some care must be taken in defining this geometric stress. For a network of Hookean springs with actual lengths $L_\alpha$ and preferred edge lengths $l_\alpha$ the energy is
\begin{align}
    E = \frac{1}{2}\sum_\alpha (L_\alpha - l_\alpha)^2.    \label{Hookean energy}
\end{align}
Following Eq~\ref{EnergyConstraints} would have us choose the constraints to be $f_\alpha = (L_\alpha - l_\alpha)$. In this case the state of self-stress is given by the tensions on the edges $\tau_\alpha = \partial E / \partial L_\alpha$, because in mechanical equilibrium we have
\begin{align}
    \frac{\partial E}{\partial x_i} = \sum_\alpha \frac{\partial E}{\partial f_\alpha}\frac{\partial f_\alpha}{\partial x_i} = \sum_\alpha \frac{\partial E}{\partial L_\alpha}R_{\alpha i} = 0.
\end{align}
However, this choice results in a nonlinear equation for the configuration that is difficult to solve. Instead, we could just as well choose $f_\alpha = L_\alpha^2/2$, which produces simpler expressions for the rigidity and prestress matrices \cite{connelly_rigidity_1980, connelly_second-order_1996, holmes-cerfon_almost-rigidity_2021}. Now the state of self-stress is not simply the tensions, but instead the force-density:
\begin{align}
    \sigma_\alpha = \frac{\partial E}{\partial f_\alpha} = \frac{1}{L_\alpha}\frac{\partial E}{\partial L_\alpha} = \frac{\tau_\alpha}{L_\alpha}. \label{force density}
\end{align}
Almost every choice of this force density can be used to solve a linear equation to obtain the corresponding configuration that possesses that state of self-stress. This property has been used by engineers to design tensioned cable structures \cite{schek_force_1974}. This means that the force-density is a geometric stress that parameterizes the critical manifold. In addition, this geometric stress is the normal vector to the critical manifold, implying that the manifold is co-dimension one~\cite{hain_optimizing_2025}. More work is needed to determine whether a similar critical manifold can be constructed in other second-order rigid underconstrained systems, such as vertex models, although their similar mathematical structure suggests it should be possible. 


\section{Beyond the limits of zero-fluctuations and linear response}
\label{sec:finite_everything}

The previous two sections focused on the mechanical response of networks in the limit of zero fluctuations and infinitely small strain. Of course, these conditions are relaxed in real biophysical systems, and we would like to understand how such perturbations affect the mechanical response. 

\subsection{Beyond linear response: plasticity, yielding, and strain stiffening}

For systems where the network can change, such as vertex or particle models, plasticity and yielding can occur if the rigid system is deformed beyond the infinitesimal strains probed by linear response. In biological networks, such large-amplitude deformations are common, and it appears that many systems have evolved to take advantage of nonlinear responses.

In these scenarios, the material is initially on the rigid side of the transition, which means that its configuration is at a metastable minima of the energy landscape, and the Hessian matrix that describes normal modes or perturbations away from that minima is positive definite.  However, finite strains will eventually push the system to a saddle point where the material becomes unstable, and then the material flows to a new metastable state.  

In first-order rigid jammed packings, the spectrum of the Hessian has a population of localized excitations at low energies, and under strain the packing undergoes localized rearrangements centered at those excitations~\cite{widmer2008irreversible, manning2011vibrational}. Depending on the preparation protocol, packings can possess a large number of such excitations and undergo ductile deformation, or possess a small number of such excitations so that rearrangements must self-organize into a sharp shear band -- an avalanche of rearrangements that occur all at the same time -- that causes brittle failure~\cite{richard2020predicting}. Researchers have developed several types of coarse-grained constitutive theories to explain this class of responses; elasto-plastic~\cite{nicolas_deformation_2018} and fluidity models~\cite{fielding2014shear} are two of the best studied options, and should also be useful for biological tissues in this regime.

For vertex models on the rigid side of the second-order rigidity transition, the density of states (e.g. distribution of the eigenvalues) and spatial organization eigenvectors in the spectrum of the Hessian are quite different from those in jammed packings~\cite{li2025connecting, schirmacher2024nature}. In addition, even on the floppy side of the transition, the material can become rigid after a finite amount of applied shear strain. This is because the vertex energy landscape is non-analytic -- it is perfectly flat with respect to small perturbations and then suddenly increases at a critical displacement of the configuration~\cite{sahu2020linear, huang_shear-driven_2021, hertaeg2024discontinuous}. Despite these important differences, the phenomenology of yielding behavior in vertex models is remarkably similar to that in jammed packings, with rearrangements occurring via localized T1 transitions~\cite{bi2014energy}. Elastoplastic models have been successfully used to characterize their behavior as well~\cite{popovic2021inferring}. 

A key open question is to what extent the yielding behavior is controlled by the rigidity transition. In jammed solids, scaling relations suggest that the yielding behavior is controlled by the underlying jamming transition~\cite{lin2014scaling}. Given the strong non-analyticity of the vertex model landscape, it is not clear that something similar holds. However, Wang and collaborators were able to quantitatively predict rearrangement rates from cell shapes in fruit fly germband extension by assuming the tissue was crossing the fluid-solid transition~\cite{wang_anisotropy_2020}, while Claussen and collaborators could quantitatively explain structure and rearrangement rates by assuming that the tissue was a solid yielding under active tensions~\cite{Shraiman2023}. Moving forward, it would be interesting to understand why both descriptions make similar predictions.

Nonlinear responses are also important in mechanical networks that do not naturally allow rearrangements, such as fiber networks. Most biological fiber networks are strain stiffening, i.e. the effective elastic modulus is much higher after the material has been strained a finite amount~\cite{sharma_strain-driven_2016}, as we discussed above as a second-order rigidity transition in spring networks, which is smoothed out by bending energy terms~\cite{licup_stress_2015}. Another class of interesting nonlinearities are those introduced by asymmetries to compression vs. extension in the response of individual springs. For example, some springs might buckle under finite compression, which can generate large-scale contraction ~\cite{lenz2012contractile}, and can even rectify active extensions on individual fibers into large-scale contraction~\cite{ronceray2016fiber}.

\subsection{Finite fluctuations:}
Another class of perturbations that are common in biological systems are thermal or actively driven fluctuations. A full discussion of how such fluctuations alter rigidity transitions is vast, so here we provide a brief review and direct interested readers towards appropriate references and reviews.

For first-order rigid systems such as jammed spheres, thermal fluctuations fundamentally alter how the system becomes rigid. Thermal disordered sphere packings undergo a glass transition~\cite{angell1995formation, StillingerGlass}, which has been discussed as one of the most difficult outstanding problems in condensed matter physics~\cite{anderson2018basic}. Practically speaking, the glass transition occurs when the dynamics of the material slow so dramatically that configurational or mechanical changes can no longer be measured~\cite{angell1995formation, StillingerGlass}.  In what follows, we will use the term ``glass transition" to generically describe the arrest of measurable flow in many different materials -- not just particle-based ones -- in the presence of general fluctuations -- not just thermal ones.

Typically, particulate glass transitions happen at densities or packing fractions $\phi_g$ that are lower than the packing fraction at which the system jams $\phi_J$, although one can transition to a glass at higher densities by cooling the system more slowly~\cite{charbonneau2017glass, parisi2010mean}. Recent computational advances suggest that 3D glasses run out of accessible configurations at a finite temperature, $T_k$, no matter how slowly they are cooled~\cite{berthier2017configurational, berthier2023modern}, which indicates that the system configuration is frozen before it undergoes a first-order rigidity transition. Understanding how this occurs is still an active area of research in statistical physics~\cite{berthier2023modern, charbonneau2017glass}.

Interestingly, different types of fluctuations (other than thermal) alter the glass transition in jammed spheres as well.  Self-propelled particles, which move persistently over a characteristic length and time scale, become rigid at a different density compared to that of a thermal glass~\cite{berthier2017active}, and recent theoretical works can explain this effect~\cite{nandi2018random}. Persistent random forces also change the yielding behavior and rheology of jammed spheres~\cite{morse_direct_2021, mandal2021shear}. See~\cite{janssen2019active} for a comprehensive review.

Second-order rigid systems also exhibit different mechanics near the transition in the presence of fluctuations. In systems where the network connectivity can change, like confluent tissues and vertex models, thermal fluctuations tend to fluidize the system, shifting the glass transition to different values of the geometric control parameter (in this case the target cell shape)~\cite{bi_motility-driven_2016}. Persistent self-propulsion also shifts the glass transition~\cite{bi_motility-driven_2016}, as does active tension fluctuations on individual edges~\cite{yamamoto2022non, devany_cell_2021}.

In second-order rigid systems with fixed network topology, such as central force networks, thermal fluctuations have the opposite effect, as they stabilize the network~\cite{zhang2016finite, arzash2023mechanical, Merkel2023}. Zhang and Mao develop an effective medium theory for disordered lattices and show that for a toy model -- similar to the stretched 3-bar-linkage (Fig.~\ref{fig:constraints}E) -- one can perform a calculation of the shear modulus using the canonical partition function $Z$. Specifically, at the second-order rigid critical point where the athermal shear modulus is zero, the linear zero modes associated with motion perpendicular to the bond give rise to a finite shear modulus $G \sim -\partial (T ln Z) / \partial x_{i\mu} \sim T^{1/2}$~\cite{zhang2016finite}. Additional numerical and theoretical work confirm this stabilization~\cite{arzash2023mechanical, Merkel2023}.

\section{Future Directions}

Although we have a foundational understanding of rigidity in biomechanical networks, major exciting open questions remain.

One class of questions involves how to interpolate between different types of models. For example, as the packing fraction increases, one might expect a crossover between particle-based first order rigidity and vertex-model-like second order rigidity. Indeed, preliminary work on deformable particle models suggests that first-order constraint counting cannot explain the onset of rigidity, and that counting second-order ``quartic modes" in the vibrational spectrum is nearly sufficient to explain rigidity~\cite{treado2021bridging}, though subtle questions remain about how to count and suggest that a full second-order analysis may be necessary.  For such systems, what are the order parameters that control the transition?

Additionally, in both vertex models and underconstrained spring network models, it is possible to increase the connectivity of the underconstrained network (in vertex models these higher coordinated vertices are called rosettes) so that the system approaches a first-order transition -- this is the red dot at $z_c$ in Fig~\ref{fig:examples}(B). Yan and Bi have developed an energetic factorization -- based on the Hessian -- to explain how this type of interpolation occurs~\cite{yan_multicellular_2019}. It would be interesting to understand more features of this mechanical interpolation -- for example, how the Hessian governs low-frequency excitations and plasticity when \emph{both} $z$ and geometric incompatibility parameters are varied.

Another class of questions involves understanding how biological systems exert robust control over rigidity transitions -- tuning themselves either towards or away from such transition points or causing transitions to occur at precise stages of development. 

One hint is that many of the biomechanical network models described above can be extended to have cell-scale properties that are dynamically tuned.  For example, active spring~\cite{staddon_mechanosensitive_2019} and tension remodeling~\cite{noll_active_2017, Shraiman2023} vertex models allow for changes in edge rest lengths and tensions. Other classes of models for fiber networks study tension-sensitive cleavage of bonds~\cite{galvani2024building} like those seen in experiments~\cite{Discher}. All of these models generate exotic, non-generic emergent mechanical states. 

Recent work in non-living systems has suggested that networks with local mechanical feedback loops on tunable degrees of freedom in the network can learn to perform tasks, termed ``physical learning"~\cite{stern2023learning, dillavou_demonstration_2022}. Physical learning principles can be used to tune rigidity in both first-order~\cite{hagh_transient_2022} and second-order~\cite{ArzashManning} rigid networks. It would be interesting to understand if these types of learning rules were operating to tune rigidity in developmental or disease states.

\section*{DISCLOSURE STATEMENT}
The authors are not aware of any affiliations, memberships, funding, or financial holdings that might be perceived as affecting the objectivity of this review. 

\section{ACKNOWLEDGMENTS}
MLM would like to thank Jane Kondev, Pankaj Mehta, and other members of the Chan Zuckerberg Theory in Biology Initiative for spurring questions about rigidity, and interest in developing this review. We also want to acknowledge useful discussions with Chris Santangelo, Miranda Cerfon-Holmes and Shlomo Gortler, and we would like to thank Alex Grigas for a critical reading of this manuscript. This project has been made possible in part by grant number 2023‐329572 from the Chan Zuckerberg Initiative DAF, an advised fund of Silicon Valley Community Foundation. MLM and TH would like to acknowledge support from NSF-DMR-1951921. KA and MLM acknowledge support from NSF-CMMI-1334611. 


\bibliographystyle{ar-style6}
\bibliography{MyLibrary}

\begin{thebibliography}{122}
\expandafter\ifx\csname natexlab\endcsname\relax\def\natexlab#1{#1}\fi

\bibitem{abbott_generalizations_2008}
Abbott TG. 2008.
Generalizations of {Kempe}'s universality theorem.
Thesis, Massachusetts Institute of Technology.
Accepted: 2009-01-30T16:38:56Z

\bibitem{alberti2021biomolecular}
Alberti S, Hyman AA. 2021.
Biomolecular condensates at the nexus of cellular stress, protein aggregation disease and ageing.
\textit{Nature reviews Molecular cell biology} 22(3):196--213

\bibitem{alshareedah2024sequence}
Alshareedah I, Borcherds WM, Cohen SR, Singh A, Posey AE, et~al. 2024.
Sequence-specific interactions determine viscoelasticity and ageing dynamics of protein condensates.
\textit{Nature Physics} 20(9):1482--1491

\bibitem{anderson2018basic}
Anderson PW. 2018.
\textit{Basic notions of condensed matter physics}.
CRC press

\bibitem{Angelini}
Angelini TE, Hannezo E, Trepat X, Marquez M, Fredberg JJ, Weitz DA. 2011{\natexlab{a}}.
Glass-like dynamics of collective cell migration.
\textit{Proceedings of the National Academy of Sciences} 108(12):4714--4719

\bibitem{angelini_glass-like_2011}
Angelini TE, Hannezo E, Trepat X, Marquez M, Fredberg JJ, Weitz DA. 2011{\natexlab{b}}.
Glass-like dynamics of collective cell migration.
\textit{Proceedings of the National Academy of Sciences} 108(12):4714--4719Publisher: Proceedings of the National Academy of Sciences

\bibitem{angell1995formation}
Angell CA. 1995.
Formation of glasses from liquids and biopolymers.
\textit{Science} 267(5206):1924--1935

\bibitem{arzash2023mechanical}
Arzash S, Gannavarapu A, MacKintosh FC. 2023.
Mechanical criticality of fiber networks at a finite temperature.
\textit{Physical Review E} 108(5):054403

\bibitem{ArzashManning}
Arzash S, Tah I, Liu AJ, Manning ML. 2023.
Tuning for fluidity using fluctuations in biological tissue models.
\textit{arXiv preprint arXiv:2312.11683}

\bibitem{benayad2020simulation}
Benayad Z, von Bulow S, Stelzl LS, Hummer G. 2020.
Simulation of fus protein condensates with an adapted coarse-grained model.
\textit{Journal of chemical theory and computation} 17(1):525--537

\bibitem{berthier2025yielding}
Berthier L, Biroli G, Manning L, Zamponi F. 2025.
Yielding and plasticity in amorphous solids.
\textit{Nature Reviews Physics} :1--18

\bibitem{berthier2017configurational}
Berthier L, Charbonneau P, Coslovich D, Ninarello A, Ozawa M, Yaida S. 2017.
Configurational entropy measurements in extremely supercooled liquids that break the glass ceiling.
\textit{Proceedings of the National Academy of Sciences} 114(43):11356--11361

\bibitem{berthier2017active}
Berthier L, Flenner E, Szamel G. 2017.
How active forces influence nonequilibrium glass transitions.
\textit{New Journal of Physics} 19(12):125006

\bibitem{berthier2023modern}
Berthier L, Reichman DR. 2023.
Modern computational studies of the glass transition.
\textit{Nature Reviews Physics} 5(2):102--116

\bibitem{bi2014energy}
Bi D, Lopez JH, Schwarz JM, Manning ML. 2014.
Energy barriers and cell migration in densely packed tissues.
\textit{Soft matter} 10(12):1885--1890

\bibitem{bi_density-independent_2015}
Bi D, Lopez JH, Schwarz JM, Manning ML. 2015{\natexlab{a}}.
A density-independent rigidity transition in biological tissues.
\textit{Nature Physics} 11(12):1074--1079Number: 12 Publisher: Nature Publishing Group

\bibitem{bi_motility-driven_2016}
Bi D, Yang X, Marchetti MC, Manning ML. 2016.
Motility-{Driven} {Glass} and {Jamming} {Transitions} in {Biological} {Tissues}.
\textit{Physical Review X} 6(2):021011Publisher: American Physical Society

\bibitem{Bi2015}
Bi DP, Lopez JH, Schwarz JM, Manning ML. 2015{\natexlab{b}}.
A density-independent rigidity transition in biological tissues.
\textit{Nature Physics} 11(12):1074--+

\bibitem{boromand_jamming_2018}
Boromand A, Signoriello A, Ye F, O’Hern CS, Shattuck MD. 2018.
Jamming of {Deformable} {Polygons}.
\textit{Physical Review Letters} 121(24):248003Publisher: American Physical Society

\bibitem{broedersz_criticality_2011}
Broedersz CP, Mao X, Lubensky TC, MacKintosh FC. 2011.
Criticality and isostaticity in fibre networks.
\textit{Nature Physics} 7(12):983--988Number: 12 Publisher: Nature Publishing Group

\bibitem{calladine_buckminster_1978}
Calladine CR. 1978.
Buckminster {Fuller}'s “{Tensegrity}” structures and {Clerk} {Maxwell}'s rules for the construction of stiff frames.
\textit{International Journal of Solids and Structures} 14(2):161--172

\bibitem{charbonneau2017glass}
Charbonneau P, Kurchan J, Parisi G, Urbani P, Zamponi F. 2017.
Glass and jamming transitions: From exact results to finite-dimensional descriptions.
\textit{Annual Review of Condensed Matter Physics} 8(1):265--288

\bibitem{chaudhuri_effects_2020}
Chaudhuri O, Cooper-White J, Janmey PA, Mooney DJ, Shenoy VB. 2020.
Effects of extracellular matrix viscoelasticity on cellular behaviour.
\textit{Nature} 584(7822):535--546Publisher: Nature Publishing Group

\bibitem{Shraiman2023}
Claussen NH, Brauns F, Shraiman BI. 2023.
A geometric tension dynamics model of epithelial convergent extension.
\textit{arXiv preprint arXiv:2311.16384}

\bibitem{connelly_rigidity_1980}
Connelly R. 1980.
The rigidity of certain cabled frameworks and the second-order rigidity of arbitrarily triangulated convex surfaces.
\textit{Advances in Mathematics} 37(3):272--299

\bibitem{connelly_higher-order_1994}
Connelly R, Servatius H. 1994.
Higher-order rigidity—{What} is the proper definition?
\textit{Discrete \& Computational Geometry} 11(2):193--200

\bibitem{connelly_second-order_1996}
Connelly R, Whiteley W. 1996.
Second-{Order} {Rigidity} and {Prestress} {Stability} for {Tensegrity} {Frameworks}.
\textit{SIAM J. Discret. Math.}

\bibitem{damavandi_energetic_2021}
Damavandi OK, Hagh VF, Santangelo CD, Manning ML. 2021.
Energetic rigidity {I}: {A} unifying theory of mechanical stability.
ArXiv:2102.11310 [cond-mat, physics:physics]

\bibitem{damavandi_energetic_2022-1}
Damavandi OK, Hagh VF, Santangelo CD, Manning ML. 2022.
Energetic rigidity. {II}. {Applications} in examples of biological and underconstrained materials.
\textit{Physical Review E} 105(2):025004Publisher: American Physical Society

\bibitem{damavandi_universal_2022}
Damavandi OK, Lawson-Keister E, Manning ML. 2022.
Universal features of rigidity transitions in vertex models for biological tissues.
Pages: 2022.06.01.494406 Section: New Results

\bibitem{StillingerGlass}
Debenedetti P, Stillinger F. 2001.
Supercooled liquids and the glass transition

\bibitem{devany_cell_2021}
Devany J, Sussman DM, Yamamoto T, Manning ML, Gardel ML. 2021.
Cell cycle-dependent active stress drives epithelia remodeling.
Pages: 804294 Section: New Results

\bibitem{dillavou_demonstration_2022}
Dillavou S, Stern M, Liu AJ, Durian DJ. 2022.
Demonstration of {Decentralized} {Physics}-{Driven} {Learning}.
\textit{Physical Review Applied} 18(1):014040Publisher: American Physical Society

\bibitem{dix2008crowding}
Dix JA, Verkman A. 2008.
Crowding effects on diffusion in solutions and cells.
\textit{Annu. Rev. Biophys.} 37(1):247--263

\bibitem{durian1995foam}
Durian DJ. 1995.
Foam mechanics at the bubble scale.
\textit{Physical review letters} 75(26):4780

\bibitem{Farhadifar}
Farhadifar R, Röper JC, Aigouy B, Eaton S, Jülicher F. 2007.
The influence of cell mechanics, cell-cell interactions, and proliferation on epithelial packing.
\textit{Current Biology} 17(24):2095--2104

\bibitem{fielding2014shear}
Fielding SM. 2014.
Shear banding in soft glassy materials.
\textit{Reports on Progress in Physics} 77(10):102601

\bibitem{fletcher_cell_2010}
Fletcher DA, Mullins RD. 2010.
Cell mechanics and the cytoskeleton.
\textit{Nature} 463(7280):485--492Publisher: Nature Publishing Group

\bibitem{galvani2024building}
Galvani~Cunha MA, Crocker JC, Liu AJ. 2024.
Building rigid networks with prestress and selective pruning.
\textit{Physical Review Research} 6(4):L042020

\bibitem{gardel2008mechanical}
Gardel ML, Kasza KE, Brangwynne CP, Liu J, Weitz DA. 2008.
Mechanical response of cytoskeletal networks.
\textit{Methods in cell biology} 89:487--519

\bibitem{gardel2006prestressed}
Gardel ML, Nakamura F, Hartwig JH, Crocker JC, Stossel TP, Weitz DA. 2006.
Prestressed f-actin networks cross-linked by hinged filamins replicate mechanical properties of cells.
\textit{Proceedings of the National Academy of Sciences} 103(6):1762--1767

\bibitem{goodrich2016scaling}
Goodrich CP, Liu AJ, Sethna JP. 2016.
Scaling ansatz for the jamming transition.
\textit{Proceedings of the National Academy of Sciences} 113(35):9745--9750

\bibitem{gortler2025higher}
Gortler SJ, Holmes-Cerfon M, Theran L. 2025.
Higher order rigidity and energy.
\textit{arXiv preprint arXiv:2506.03108}

\bibitem{gottheil2023state}
Gottheil P, Lippoldt J, Grosser S, Renner F, Saibah M, et~al. 2023.
State of cell unjamming correlates with distant metastasis in cancer patients.
\textit{Physical Review X} 13(3):031003

\bibitem{grigas2025protein}
Grigas AT, Liu Z, Logan JA, Shattuck MD, O'Hern CS. 2025.
Protein folding as a jamming transition.
\textit{PRX Life} 3(1):013018

\bibitem{grosser2021cell}
Grosser S, Lippoldt J, Oswald L, Merkel M, Sussman DM, et~al. 2021.
Cell and nucleus shape as an indicator of tissue fluidity in carcinoma.
\textit{Physical Review X} 11(1):011033

\bibitem{hagh_transient_2022}
Hagh VF, Nagel SR, Liu AJ, Manning ML, Corwin EI. 2022.
Transient learning degrees of freedom for introducing function in materials.
\textit{Proceedings of the National Academy of Sciences} 119(19):e2117622119Publisher: Proceedings of the National Academy of Sciences

\bibitem{hain_optimizing_2025}
Hain T, Santangelo C, Manning ML. 2025.
Optimizing properties on the critical rigidity manifold of underconstrained central-force networks.
\textit{Physical Review E} 111(1):015418Publisher: American Physical Society

\bibitem{Shenoy2}
Hall MS, Alisafaei F, Ban E, Feng X, Hui CY, et~al. 2016.
Fibrous nonlinear elasticity enables positive mechanical feedback between cells and ecms.
\textit{Proceedings of the National Academy of Sciences} 113(49):14043--14048

\bibitem{head_distinct_2003}
Head DA, Levine AJ, MacKintosh FC. 2003.
Distinct regimes of elastic response and deformation modes of cross-linked cytoskeletal and semiflexible polymer networks.
\textit{Physical Review E} 68(6):061907Publisher: American Physical Society

\bibitem{hertaeg2024discontinuous}
Hertaeg MJ, Fielding SM, Bi D. 2024.
Discontinuous shear thickening in biological tissue rheology.
\textit{Physical Review X} 14(1):011027

\bibitem{holmes-cerfon_almost-rigidity_2021}
Holmes-Cerfon M, Theran L, Gortler SJ. 2021.
Almost-{Rigidity} of {Frameworks}.
\textit{Communications on Pure and Applied Mathematics} 74(10):2185--2247\_eprint: https://onlinelibrary.wiley.com/doi/pdf/10.1002/cpa.21971

\bibitem{honda_three-dimensional_2004}
Honda H, Tanemura M, Nagai T. 2004.
A three-dimensional vertex dynamics cell model of space-filling polyhedra simulating cell behavior in a cell aggregate.
\textit{Journal of Theoretical Biology} 226(4):439--453

\bibitem{huang_shear-driven_2021}
Huang J, Cochran JO, Fielding SM, Marchetti MC, Bi D. 2021.
Shear-driven solidification and nonlinear elasticity in epithelial tissues.
\textit{arXiv:2109.10374 [cond-mat, physics:physics, q-bio]} ArXiv: 2109.10374

\bibitem{jansen2018role}
Jansen KA, Licup AJ, Sharma A, Rens R, MacKintosh FC, Koenderink GH. 2018.
The role of network architecture in collagen mechanics.
\textit{Biophysical journal} 114(11):2665--2678

\bibitem{janssen2019active}
Janssen LM. 2019.
Active glasses.
\textit{Journal of Physics: Condensed Matter} 31(50):503002

\bibitem{jawerth2020protein}
Jawerth L, Fischer-Friedrich E, Saha S, Wang J, Franzmann T, et~al. 2020.
Protein condensates as aging maxwell fluids.
\textit{Science} 370(6522):1317--1323

\bibitem{kang2015confinement}
Kang H, Yoon YG, Thirumalai D, Hyeon C. 2015.
Confinement-induced glassy dynamics in a model for chromosome organization.
\textit{Physical review letters} 115(19):198102

\bibitem{ActiveFoam}
Kim S, Pochitaloff M, Stooke-Vaughan GA, Campàs O. 2021.
Embryonic tissues as active foams.
\textit{Nature Physics} 17(7):859--866

\bibitem{koeze2018sticky}
Koeze DJ, Tighe BP. 2018.
Sticky matters: Jamming and rigid cluster statistics with attractive particle interactions.
\textit{Physical review letters} 121(18):188002

\bibitem{Merkel2023}
Lee CT, Merkel M. 2023.
Generic elasticity of thermal, under-constrained systems.
\textit{arXiv preprint arXiv:2304.07266}

\bibitem{lenz2012contractile}
Lenz M, Thoresen T, Gardel ML, Dinner AR. 2012.
Contractile units in disordered actomyosin bundles arise from f-actin buckling.
\textit{Physical review letters} 108(23):238107

\bibitem{li2025connecting}
Li C, Merkel M, Sussman DM. 2025.
Connecting anomalous elasticity and sub-arrhenius structural dynamics in a cell-based model.
\textit{Physical Review Letters} 134(4):048203

\bibitem{licup_stress_2015}
Licup A, Münster S, Sharma A, Sheinman M, Jawerth L, et~al. 2015.
Stress controls the mechanics of collagen networks.
\textit{Proceedings of the National Academy of Sciences of the United States of America} 112(31):9573--9578

\bibitem{lin2014scaling}
Lin J, Lerner E, Rosso A, Wyart M. 2014.
Scaling description of the yielding transition in soft amorphous solids at zero temperature.
\textit{Proceedings of the National Academy of Sciences} 111(40):14382--14387

\bibitem{loewe_solid-liquid_2020}
Loewe B, Chiang M, Marenduzzo D, Marchetti MC. 2020.
Solid-{Liquid} {Transition} of {Deformable} and {Overlapping} {Active} {Particles}.
\textit{Physical Review Letters} 125(3):038003Publisher: American Physical Society

\bibitem{lubensky_phonons_2015}
Lubensky TC, Kane CL, Mao X, Souslov A, Sun K. 2015.
Phonons and elasticity in critically coordinated lattices.
\textit{arXiv:1503.01324 [cond-mat]} ArXiv: 1503.01324

\bibitem{mackintosh_elasticity_1995}
MacKintosh FC, Käs J, Janmey PA. 1995.
Elasticity of {Semiflexible} {Biopolymer} {Networks}.
\textit{Physical Review Letters} 75(24):4425--4428Publisher: American Physical Society

\bibitem{mandal2021shear}
Mandal R, Sollich P. 2021.
Shear-induced orientational ordering in an active glass former.
\textit{Proceedings of the National Academy of Sciences} 118(39):e2101964118

\bibitem{manning2011vibrational}
Manning ML, Liu AJ. 2011.
Vibrational modes identify soft spots in a sheared disordered packing.
\textit{Physical Review Letters} 107(10):108302

\bibitem{maxwell_l_1864}
Maxwell JC. 1864.
L. {On} the calculation of the equilibrium and stiffness of frames.
\textit{The London, Edinburgh, and Dublin Philosophical Magazine and Journal of Science} 27(182):294--299Publisher: Taylor \& Francis \_eprint: https://doi.org/10.1080/14786446408643668

\bibitem{mcguffee2010diffusion}
McGuffee SR, Elcock AH. 2010.
Diffusion, crowding \& protein stability in a dynamic molecular model of the bacterial cytoplasm.
\textit{PLoS computational biology} 6(3):e1000694

\bibitem{merkel_minimal-length_2019}
Merkel M, Baumgarten K, Tighe BP, Manning ML. 2019.
A minimal-length approach unifies rigidity in underconstrained materials.
\textit{Proceedings of the National Academy of Sciences} 116(14):6560--6568Publisher: Proceedings of the National Academy of Sciences

\bibitem{merkel_geometrically_2018}
Merkel M, Manning ML. 2018.
A geometrically controlled rigidity transition in a model for confluent {3D} tissues.
\textit{New Journal of Physics} 20(2):022002Publisher: IOP Publishing

\bibitem{michieletto2016polymer}
Michieletto D, Orlandini E, Marenduzzo D. 2016.
Polymer model with epigenetic recoloring reveals a pathway for the de novo establishment and 3d organization of chromatin domains.
\textit{Physical Review X} 6(4):041047

\bibitem{miroshnikova2018adhesion}
Miroshnikova YA, Le HQ, Schneider D, Thalheim T, R{\"u}bsam M, et~al. 2018.
Adhesion forces and cortical tension couple cell proliferation and differentiation to drive epidermal stratification.
\textit{Nature cell biology} 20(1):69--80

\bibitem{mizuno_nonequilibrium_2007}
Mizuno D, Tardin C, Schmidt CF, MacKintosh FC. 2007.
Nonequilibrium {Mechanics} of {Active} {Cytoskeletal} {Networks}.
\textit{Science} 315(5810):370--373Publisher: American Association for the Advancement of Science

\bibitem{MongeraCampas}
Mongera A, Rowghanian P, Gustafson HJ, Shelton E, Kealhofer DA, et~al. 2018{\natexlab{a}}.
A fluid-to-solid jamming transition underlies vertebrate body axis elongation.
\textit{Nature} :1

\bibitem{mongera_fluid--solid_2018}
Mongera A, Rowghanian P, Gustafson HJ, Shelton E, Kealhofer DA, et~al. 2018{\natexlab{b}}.
A fluid-to-solid jamming transition underlies vertebrate body axis elongation.
\textit{Nature} 561(7723):401--405Number: 7723 Publisher: Nature Publishing Group

\bibitem{morse_direct_2021}
Morse PK, Roy S, Agoritsas E, Stanifer E, Corwin EI, Manning ML. 2021.
A direct link between active matter and sheared granular systems.
\textit{Proceedings of the National Academy of Sciences} 118(18):e2019909118

\bibitem{moshe2018}
Moshe M, Bowick MJ, Marchetti MC. 2018.
Geometric frustration and solid-solid transitions in model 2d tissue.
\textit{Physical review letters} 120(26):268105

\bibitem{muiznieks_molecular_2013}
Muiznieks LD, Keeley FW. 2013.
Molecular assembly and mechanical properties of the extracellular matrix: {A} fibrous protein perspective.
\textit{Biochimica et Biophysica Acta (BBA) - Molecular Basis of Disease} 1832(7):866--875

\bibitem{muller2015marginal}
M{\"u}ller M, Wyart M. 2015.
Marginal stability in structural, spin, and electron glasses.
\textit{Annu. Rev. Condens. Matter Phys.} 6(1):177--200

\bibitem{nagai_dynamic_2001}
Nagai T, Honda H. 2001.
A dynamic cell model for the formation of epithelial tissue.
\textit{Philosophical Magazine B} 81:699--719

\bibitem{nandi2018random}
Nandi SK, Mandal R, Bhuyan PJ, Dasgupta C, Rao M, Gov NS. 2018.
A random first-order transition theory for an active glass.
\textit{Proceedings of the National Academy of Sciences} 115(30):7688--7693

\bibitem{nicolas_deformation_2018}
Nicolas A, Ferrero EE, Martens K, Barrat JL. 2018.
Deformation and flow of amorphous solids: {Insights} from elastoplastic models.
\textit{Reviews of Modern Physics} 90(4):045006Publisher: American Physical Society

\bibitem{noll_active_2017}
Noll N, Mani M, Heemskerk I, Streichan SJ, Shraiman BI. 2017.
Active {Tension} {Network} model suggests an exotic mechanical state realized in epithelial tissues.
\textit{Nature Physics} 13(12):1221--1226

\bibitem{nonomura2012study}
Nonomura M. 2012.
Study on multicellular systems using a phase field model.
\textit{PloS one} 7(4):e33501

\bibitem{damavandi_universal}
Ojan~Khatib D, Elizabeth LK, Manning ML. 2022.
Universal features of rigidity transitions in vertex models for biological tissues.
\textit{bioRxiv} :2022.06.01.494406

\bibitem{KasCancerJamming}
Oswald L, Grosser S, Smith DM, Käs JA. 2017{\natexlab{a}}.
Jamming transitions in cancer.
\textit{Journal of Physics D: Applied Physics} 50(48):483001

\bibitem{oswald_jamming_2017}
Oswald L, Grosser S, Smith DM, Käs JA. 2017{\natexlab{b}}.
Jamming transitions in cancer.
\textit{Journal of Physics D: Applied Physics} 50(48):483001Publisher: IOP Publishing

\bibitem{o2003jamming}
O’hern CS, Silbert LE, Liu AJ, Nagel SR. 2003.
Jamming at zero temperature and zero applied stress: The epitome of disorder.
\textit{Physical Review E} 68(1):011306

\bibitem{parisi2010mean}
Parisi G, Zamponi F. 2010.
Mean-field theory of hard sphere glasses and jamming.
\textit{Reviews of Modern Physics} 82(1):789--845

\bibitem{park_unjamming_2015}
Park JA, Kim JH, Bi D, Mitchel JA, Qazvini NT, et~al. 2015.
Unjamming and cell shape in the asthmatic airway epithelium.
\textit{Nature Materials} 14(10):1040--1048Number: 10 Publisher: Nature Publishing Group

\bibitem{petridou_rigidity_2021}
Petridou NI, Corominas-Murtra B, Heisenberg CP, Hannezo E. 2021.
Rigidity percolation uncovers a structural basis for embryonic tissue phase transitions.
\textit{Cell} 184(7):1914--1928.e19

\bibitem{popovic2021inferring}
Popovi{\'c} M, Druelle V, Dye NA, J{\"u}licher F, Wyart M. 2021.
Inferring the flow properties of epithelial tissues from their geometry.
\textit{New Journal of Physics} 23(3):033004

\bibitem{pritchard2014mechanics}
Pritchard RH, Huang YYS, Terentjev EM. 2014.
Mechanics of biological networks: from the cell cytoskeleton to connective tissue.
\textit{Soft matter} 10(12):1864--1884

\bibitem{pujol2012impact}
Pujol T, Du~Roure O, Fermigier M, Heuvingh J. 2012.
Impact of branching on the elasticity of actin networks.
\textit{Proceedings of the National Academy of Sciences} 109(26):10364--10369

\bibitem{richard2020predicting}
Richard D, Ozawa M, Patinet S, Stanifer E, Shang B, et~al. 2020.
Predicting plasticity in disordered solids from structural indicators.
\textit{Physical Review Materials} 4(11):113609

\bibitem{ronceray2016fiber}
Ronceray P, Broedersz CP, Lenz M. 2016.
Fiber networks amplify active stress.
\textit{Proceedings of the national academy of sciences} 113(11):2827--2832

\bibitem{sahu2020linear}
Sahu P, Kang J, Erdemci-Tandogan G, Manning ML. 2020.
Linear and nonlinear mechanical responses can be quite different in models for biological tissues.
\textit{Soft Matter} 16(7):1850--1856

\bibitem{Discher}
Saini K, Cho S, Dooling LJ, Discher DE. 2020.
Tension in fibrils suppresses their enzymatic degradation–a molecular mechanism for ‘use it or lose it’.
\textit{Matrix Biology} 85:34--46

\bibitem{schek_force_1974}
Schek HJ. 1974.
The force density method for form finding and computation of general networks.
\textit{Computer Methods in Applied Mechanics and Engineering} 3(1):115--134

\bibitem{schenk_zero_2014}
Schenk M, Guest SD. 2014.
On zero stiffness.
\textit{Proceedings of the Institution of Mechanical Engineers, Part C: Journal of Mechanical Engineering Science} 228(10):1701--1714Publisher: IMECHE

\bibitem{schirmacher2024nature}
Schirmacher W, Paoluzzi M, Mocanu FC, Khomenko D, Szamel G, et~al. 2024.
The nature of non-phononic excitations in disordered systems.
\textit{Nature Communications} 15(1):3107

\bibitem{serwane_vivo_2017}
Serwane F, Mongera A, Rowghanian P, Kealhofer DA, Lucio AA, et~al. 2017.
In vivo quantification of spatially varying mechanical properties in developing tissues.
\textit{Nature Methods} 14(2):181--186Publisher: Nature Publishing Group

\bibitem{sharma_strain-controlled_2016}
Sharma A, Licup AJ, Jansen KA, Rens R, Sheinman M, et~al. 2016{\natexlab{a}}.
Strain-controlled criticality governs the nonlinear mechanics of fibre networks.
\textit{Nature Physics} 12(6):584--587Number: 6 Publisher: Nature Publishing Group

\bibitem{sharma_strain-driven_2016}
Sharma A, Licup AJ, Rens R, Vahabi M, Jansen KA, et~al. 2016{\natexlab{b}}.
Strain-driven criticality underlies nonlinear mechanics of fibrous networks.
\textit{Physical Review E} 94(4):042407Publisher: American Physical Society

\bibitem{shen2023liquid}
Shen Y, Chen A, Wang W, Shen Y, Ruggeri FS, et~al. 2023.
The liquid-to-solid transition of fus is promoted by the condensate surface.
\textit{Proceedings of the National Academy of Sciences} 120(33):e2301366120

\bibitem{staddon_mechanosensitive_2019}
Staddon MF, Cavanaugh KE, Munro EM, Gardel ML, Banerjee S. 2019.
Mechanosensitive {Junction} {Remodeling} {Promotes} {Robust} {Epithelial} {Morphogenesis}.
\textit{Biophysical Journal} 117(9):1739--1750

\bibitem{stern2023learning}
Stern M, Murugan A. 2023.
Learning without neurons in physical systems.
\textit{Annual Review of Condensed Matter Physics} 14(1):417--441

\bibitem{treado2021bridging}
Treado JD, Wang D, Boromand A, Murrell MP, Shattuck MD, O'Hern CS. 2021.
Bridging particle deformability and collective response in soft solids.
\textit{Physical Review Materials} 5(5):055605

\bibitem{van2009leveraging}
Van~Mameren J, Vermeulen KC, Gittes F, Schmidt CF. 2009.
Leveraging single protein polymers to measure flexural rigidity.
\textit{The Journal of Physical Chemistry B} 113(12):3837--3844

\bibitem{vermeulen_geometry_2017}
Vermeulen MFJ, Bose A, Storm C, Ellenbroek WG. 2017.
Geometry and the onset of rigidity in a disordered network.
\textit{Physical Review E} 96(5):053003Publisher: American Physical Society

\bibitem{wang_anisotropy_2020}
Wang X, Merkel M, Sutter LB, Erdemci-Tandogan G, Manning ML, Kasza KE. 2020.
Anisotropy links cell shapes to tissue flow during convergent extension.
\textit{Proceedings of the National Academy of Sciences} 117(24):13541--13551Publisher: Proceedings of the National Academy of Sciences

\bibitem{Weaire}
Weaire D, Hutzler S. 2001.
\textit{The physics of foams}.
Oxford: Oxford University Press

\bibitem{widmer2008irreversible}
Widmer-Cooper A, Perry H, Harrowell P, Reichman DR. 2008.
Irreversible reorganization in a supercooled liquid originates from localized soft modes.
\textit{Nature Physics} 4(9):711--715

\bibitem{wu2022vimentin}
Wu H, Shen Y, Sivagurunathan S, Weber MS, Adam SA, et~al. 2022.
Vimentin intermediate filaments and filamentous actin form unexpected interpenetrating networks that redefine the cell cortex.
\textit{Proceedings of the National Academy of Sciences} 119(10):e2115217119

\bibitem{wyse2022structural}
Wyse~Jackson T, Michel J, Lwin P, Fortier LA, Das M, et~al. 2022.
Structural origins of cartilage shear mechanics.
\textit{Science Advances} 8(6):eabk2805

\bibitem{yamamoto2022non}
Yamamoto T, Sussman DM, Shibata T, Manning ML. 2022.
Non-monotonic fluidization generated by fluctuating edge tensions in confluent tissues.
\textit{Soft Matter} 18(11):2168--2175

\bibitem{yan_multicellular_2019}
Yan L, Bi D. 2019.
Multicellular {Rosettes} {Drive} {Fluid}-solid {Transition} in {Epithelial} {Tissues}.
\textit{Physical Review X} 9(1):011029Publisher: American Physical Society

\bibitem{zhang2016finite}
Zhang L, Mao X. 2016.
Finite-temperature mechanical instability in disordered lattices.
\textit{Physical Review E} 93(2):022110

\end{thebibliography}
\end{document}